# Undocumented Behavior in the `gsynth` R package and its Consequences for Three Published Studies

Initial manuscript date: April 13, 2026. Current manuscript date: September 16, 2026.

Beniamino Green beniamino.green@yale.edu
P.M. Aronow p.aronow@yale.edu

**Abstract**: Prior to the version 1.3.1 update on CRAN in December 2025, `gsynth`, a popular R package for estimating Interactive Fixed Effects (IFE) models, could drastically and systematically underestimate standard errors. This underestimation would occur when two estimation options (`inference = "parametric"`, and `EM = TRUE`) were used together, in which case the package would apply a parametric bootstrap procedure to Gobillon and Magnac (2016)'s IFE-EM estimator. The package ceased supporting this combination in December 2025, and the latest documentation now describes the parametric bootstrap as not suitable for use with the IFE-EM estimator due to a theoretical incompatibility. Our focus is an implementation error we identified in the pre-1.3.1 versions of `gsynth`: the parametric bootstrap used when `EM = TRUE` did not match the algorithm proposed in Xu (2017), using in-sample residuals instead of out-of-sample errors. We show that this implementation error can cause underestimation by orders of magnitude. We conduct an empirical Monte Carlo study using randomly assigned placebo treatments on a series of state-level panel datasets, and show that `gsynth` could yield high false positive rates in realistic settings. We identify three papers published in the *American Political Science Review* that are affected by this behavior. Reanalyzing the relevant sections of these papers, we show that (i) correcting the implementation error renders most findings insignificant, and (ii) using Xu (2017)'s Generalized Synthetic Control method in place of IFE-EM renders every finding insignificant.

Replication materials available at: https://github.com/beniaminogreen/green_aronow_2026_replication

**UPDATE** (APRIL 28, 2026):

We shared our manuscript with Xu who posted a response acknowledging the implementation error documented here. We thank Xu for his extended engagement and prompt response to our concerns. From its first CRAN release until December 2025, `gsynth` allowed parametric inference under IFE-EM using an undocumented routine that omitted the out-of-sample leave-one-out prediction-error step prescribed in Xu (2017) for GSC. Xu agrees that this error can lead to severe undercoverage under serial correlation and overfitting. He also argues that simply restoring the leave-one-out step does not itself provide a valid parametric bootstrap for IFE-EM, because IFE-EM estimates the factor space using treated units' pre-treatment outcomes, whereas GSC estimates the factor space from never-treated controls only. We therefore use "corrected" below in an implementation-specific sense: our IFE-EM correction restores the omitted step from the published Xu (2017) algorithm, and is useful for isolating the consequences of the historical implementation error, but it should not be read as an endorsement of parametric-bootstrap inference for IFE-EM.

Xu's reanalyses of the three APSR applications broadly support our conclusion that the affected analyses were overconfident, although his updated diagnostics and package implementations produce a small number of exceptions under alternative rank choices. We leave our benchmark reanalyses intact because they evaluate the original specifications and the historical code path used by the published applications, but Xu's response reinforces the practical recommendation that applied users report rank sensitivity and placebo diagnostics, distinguish GSC from IFE-EM, and use an inferential procedure with theoretical support for the estimator actually being fit. We do not otherwise speak to the theoretical or empirical claims in Xu's response.

**UPDATE** (JUNE 8, 2026):

After receiving comments on the paper, including from the authors of the papers we replicate, we made the following two revisions to improve the clarity of the manuscript:

- We updated description of the analysis in Alsaadi (2025) to give readers more context on how the GSC models we examine fit into the paper; the original version referred to the GSC models simply 'as part of a larger analysis'. We thank Salam Alsaadi for this suggestion.
- We updated the introduction to the *Interactive Fixed Effects Models in Gsynth* section to specifically identify that the bug we discuss involves using in-sample residuals instead of prediction errors - previous versions referred to the difference between the proposed and implemented bootstrap procedures as simply 'a discrepancy'.

**UPDATE** (JUNE 17, 2026):

We note that updates have recently been made to the `gsynth` documentation, and thus our statements about the state of the package may no longer be current. We do not intend to update this manuscript as a result of these or any further changes to `gsynth`.

We made three additional revisions:

- We added a link to the replication archive (https://github.com/beniaminogreen/green_aronow_2026_replication) on the title page.
- We corrected a spelling error in the June 8th update.
- We added a declaration of generative AI and AI-assisted technologies on page 15.

**Update** (September 16, 2026):

We conducted an audit of this work using large language models, which reported issues that we then reviewed. We updated the declaration of Generative AI accordingly.

First, we discovered that two additional undocumented behaviors in the `gsynth` 1.2.1 GSC bootstrap were passed through to our corrected IFE-EM bootstrap and the analysis reported here. In the final bootstrap, the code resamples control identities with replacement, carrying their fitted outcomes, covariates, and observation patterns into each draw (`R/core.R`: gsynth, lines 2600–2647 and our correction, lines 2879–2926). Step 3 of the published pseudocode in Xu (2017, Algorithm 2, pp. 65–66) instead adds a drawn residual to each original control's fitted outcome; its control resampling occurs in Step 1, when constructing prediction errors. The code also repeats one selected control as many times as there are treated units, assigning the copies the original treated units' treatment patterns (`R/core.R`: gsynth, lines 2494–2529 and our correction, lines 2768–2804). Step 1 of the pseudocode instead forms a sample with one pseudo-treated unit and resampled controls. We found no prose explanation of either behavior in the contemporary materials searched. We leave both unchanged and do not investigate them further. Specifically, we have not closely investigated whether or not these undocumented behaviors still persist in `gsynth`'s successor package, `fect`. We have narrowed our description of the correction to restoration of the omitted prediction-error step, rather than agreement with the full published algorithm. We removed the word `alone` from the abstract, when attributing issues to the implementation error.

Second, we found errors in our reporting of our results, which we enumerate below. All code, figures, and numerical results are retained. We also fixed some minor copyediting errors, and added more general language about how to search for affected code to the guide for affected papers.

- **Toy-example parameters.** We erroneously reported parameters used to generate our 'toy example'. Previously, we reported using variance 10.2 and covariance $10e^{-\frac{(t-t')^2}{600}}$. We corrected these to reflect the values used in the simulation: variance 4.01 and covariance $4e^{-\frac{(t-t')^2}{300}}$ for distinct periods.
- **Off-by-One Errors:** We misreported the number of pre-treatment periods and units by one across several simulations, specifically:
  - In our toy example simulation, we reported that treatment started at $t = 40$ rather than $t = 41$, which was the correct value.
  - Our empirical Monte Carlo panels contained 19 periods, rather than the reported 20. Treatment began in period 12, leaving 11 pretreatment periods and eight treatment periods.
- **Dataset Sizes**:
  - Originally, we reported that our Monte Carlo studies used 50 units. In fact, six panels contain 51 units, including the District of Columbia, rather than only the 50 states: population, income, GDP per capita, the Gini coefficient, violent crime, and policy mood. The other ten panels contain 50 states.
  - For Eibl and Hertog (2023), we added text distinguishing the outcome data's 1900-2018 span (119 years) from the fitted panels' 61–65 periods and 73–152 countries across the main and appendix analyses.
  - For Alsaadi (2025), we now report the span of the data for the fitted panel, which runs from 1900-2015, rather than the previously-reported 1885-2015. The number of countries used in the analysis is unchanged.
- **Corporate-tax Rate Effects** The previous version of this paper described estimated 4 and 8 *percentage-point* reductions in corporate tax rates. We corrected this to reflect that these estimated reductions are *percentages*, not *percentage points*. We also clarify that our reported sample sizes apply specifically to the analyses of corporate tax rates, not other legislative outcomes.

- **Unbalanced-panel bootstrap errors.** We clarified that empirical error trajectories are resampled only for complete panels. Unbalanced panels use mean-zero multivariate Normal draws. Historical IFE-EM generates errors for both treated and control units using a second-moment matrix estimated from control-unit residuals ; GSC and our corrected bootstrap use prediction-error matrices for treated units and residual matrices for controls.

# Introduction

Interactive Fixed Effects (IFE) models are often used for the analysis of panel data, and relax the conventional parallel-trends assumption needed to estimate causal effects. Since their introduction to the field of political science by Xu (2017), these methods have surged in popularity, being used to study a diverse set of phenomena including the political effects of Supreme Court decisions (Gilens, Patterson, and Haines 2021), the provision of public goods in oil-rich states (Eibl and Hertog 2023), and democratic backsliding (Grumbach 2022). Outside of political science, IFE models have seen a similar pattern of adoption, being used to analyze the effects of carbon offsets (Tang et al. 2025), vaccine mandates (Mills and Rüttenauer 2022), and congestion pricing (Cook et al. 2025), inter alia.

IFE models are often estimated using the `gsynth` package for R, which receives over 300 downloads a week as of April 2026. In this note, we show that before version 1.3.1 on CRAN, when two options were used in tandem (the parametric bootstrap and IFE-EM), the resulting standard errors could be severely underestimated. This issue could arise from two sources. First, prior to version 1.3.1, `gsynth` would allow users to combine two options, the IFE-EM estimator (Gobillon and Magnac 2016) and a parametric bootstrap (Xu 2017) for inference. The parametric bootstrap was recommended when the number of treated units was small, and IFE-EM was recommended for potential efficiency gains. A package vignette demonstrated using these settings in tandem, suggesting that the two approaches could be used together. However, the current package documentation asserts a theoretical incompatibility between IFE-EM and the parametric bootstrap, and no longer recommends their joint use.

Second, we discovered an implementation error in the parametric bootstrap procedure that is called when IFE-EM is selected in all versions published to CRAN prior to 1.3.1.[1] This parametric bootstrap procedure does not match the algorithm proposed in Xu (2017). Specifically, this procedure generates new samples using only residuals instead of a combination of residuals and out-of-sample prediction errors, which mechanically decreases the variance of estimates across bootstrapped samples. This results in smaller standard errors, often by orders of magnitude. Setting aside any theoretical questions about the suitability of Xu (2017)'s parametric bootstrap, this implementation error alone is sufficient to materially understate uncertainty.

We show that the package's historical behavior has been consequential for published work, including three papers in the *American Political Science Review* (APSR). We report three inferential procedures: (i) the original results, using IFE-EM with the `gsynth` 1.2.1 implementation of the parametric bootstrap; (ii) IFE-EM with a corrected version of the parametric bootstrap that restores the omitted prediction-error step from Xu (2017)'s published algorithm; and (iii) the Generalized Synthetic Control (GSC) method as described in the associated paper (Xu 2017) and as currently recommended by the `gsynth` package. We demonstrate that use of the corrected parametric bootstrap leads almost all findings to become insignificant at conventional levels. Furthermore, use of the GSC method leads all results to become insignificant.

# Interactive Fixed Effects Models in `gsynth`

In this section, we explore how the `gsynth` package estimates causal effects and confidence intervals. We highlight the procedure used when `inference='parametric'` and `EM=TRUE` are set, and we discuss how this procedure can lead to undercoverage due to a discrepancy between the implementation and the procedure described in the paper (Xu 2017). Specifically, the parametric bootstrap procedure proposed by Xu (2017) generates bootstrap samples by resampling out-of-sample prediction errors and adding these to the fitted values from the GSC model. When `inference = "parametric"` and `EM=T` are used together, the bootstrap procedure uses in-sample residuals in place of out-of-sample errors, which can lead to drastic under-coverage. We note that the `EM=TRUE` parameter corresponds to using the IFE-EM estimator (Gobillon and Magnac 2016), which uses a different fitting approach than the GSC method described in Xu (2017).

[1] As of April 2026, this error is now acknowledged in the `gsynth` documentation: "If your existing code uses estimator = "ife" with EM = TRUE, which maps to IFE-EM, do not use inference = "parametric"; a previous version contained an erroneous implementation, which severely underestimates uncertainties. It was removed in v1.3.1."

### Treatment Effect Estimation

First, we fix notation, largely following that of Xu (2017). Assume that there exist $N$ units indexed $i \in \{1, ..., N\}$ which are observed for $T$ time periods indexed $t \in \{1, ..., T\}$. Each unit has a pair of potential outcomes $\left(Y_{i,t}(0), Y_{i,t}(1)\right)$ which are revealed depending on the unit's treatment status at a given time period. Let $\mathbb{T}$ denote the set of treated observations such that $\mathbb{T} = \{(i, t) \mid \text{unit } i \text{ is treated at time } t\}$, and denote the sets of treated and untreated units as $\mathcal{T}$ and $\mathcal{C}$.

The estimand of interest is the average treatment effect on the treated units (ATT), which can be written as:

$$\text{ATT} = \frac{1}{|\mathbb{T}|} \sum_{(i,t) \in \mathbb{T}} Y_{i,t}(1) - Y_{i,t}(0)$$

The ATT is estimated by imputing counterfactuals $Y_{i,t}(0)$ for units under treatment, assuming the following generative model for control observations:

$$Y_{\text{it}}(0) = \underbrace{x_{\text{it}}^T \beta}_{\text{covariate adjustment}} + \underbrace{\lambda_i^T f_t}_{\text{latent factors}} + \underbrace{\gamma_t + \eta_i}_{\substack{\text{fixed effects} \\ \text{(optional)}}} + \underbrace{\varepsilon_{\text{it}}}_{\text{shocks}}$$

If fixed effects are used, then the model is identical to a classical two-way fixed-effects model apart from the terms $\lambda$ and $f$ which allow the model to adjust for the non-parallel evolution of the treated and control groups. The terms $f$ are latent factors which are shared between all units and evolve over time, while the terms $\lambda$ are 'loadings' which are unique to each unit and constant over time. The interaction of the latent factors and loadings gives the IFE model its name, and allows the method to adjust for complex, unit-specific trends that can be expressed as a combination of time-varying confounding factors.

Importantly, the number of factors to estimate is not known *a priori* and must be estimated from the data. This choice is consequential in determining the coverage properties of the estimator. The number of factors is chosen via a cross-validation step that minimizes the leave-one-out prediction error across held-out pre-treatment time periods. Specifically, the procedure works as follows: for each possible choice of $k$, estimate $k$ latent factors, holding back one time period for validation. The $k$ that obtains the smallest prediction error across held-out time periods is chosen as the number of factors used to impute counterfactual outcomes.

### GSC and IFE-EM

The `gsynth` package provides two ways to fit interactive fixed-effects models which differ in how they use pre-treatment observations. The Generalized Synthetic Control (GSC) method estimates latent factors using only never-treated units and then estimates the loadings for treated units by regressing pre-treatment outcomes on the estimated factors. This procedure allows for consistent estimates of causal effects, but leads to a possible efficiency loss as it does not make full use of pre-treatment data. In earlier work, Gobillon and Magnac (2016) developed a similar estimation algorithm that uses an expectation-maximization step to incorporate pre-treatment observations when estimating the latent factors. We follow recent versions of the `gsynth` documentation, and refer to this algorithm as the IFE-EM algorithm. In summary:

- The IFE-EM estimator (Gobillon and Magnac 2016) fits latent factors based on *both* never-treated and not-yet-treated units, jointly estimating factors and loadings for all units.
- The GSC estimator (Xu 2017) estimates latent factors using *only* never-treated units, then regresses the pre-treatment outcomes from treated units on these estimated factors.

The `gsynth` package implements both the GSC estimation routine and the IFE-EM algorithm, but has historically elided the difference between the two methods; in prior versions of the `gsynth` package, both estimation routines were called `ife`, and users chose between the two estimators by selecting `EM=F` (in which case the GSC estimator was used), or `EM=T` (in which case IFE-EM was used). This led to a situation in which all uses of IFE-EM were referred to as GSC or GSC+EM by applied researchers.

This terminology may make it difficult to conclude from the text of a paper whether GSC or IFE-EM was used, which is important for understanding whether the confidence intervals are vulnerable to the issue we highlight. We note that all of the APSR papers we review in our reanalysis that use IFE-EM refer to it as the GSC method without qualification.

### The Parametric Bootstrap Procedure for GSC

The original GSC paper describes a parametric bootstrap approach tailored for the small samples typically encountered in political science (Xu 2017, p. 64). For complete panels, the procedure works by resampling residuals and prediction errors to produce new datasets in a manner similar to a residual bootstrap. Observations in each bootstrapped sample are generated by taking the predicted outcomes from the factor model, and adding a resampled residual or out-of-sample prediction error. This process is described in detail in Algorithm 2 of the paper introducing the method (Xu 2017, p. 66), which we reproduce below:

Bootstrap Procedure (GSC), Reproduced from Xu (2017)

**Step 1. Start** a loop that runs $B_1$ times:

In round $m \in \{1, ..., B_1\}$, randomly select one control unit $i$ as if it was treated when $t > T_0$.

Resample the rest of the control group with replacement of size $N_{\text{co}}$ and form a new sample with one "treated" unit and $N_{\text{co}}$ resampled control units.

Apply the GSC method to the new sample, obtaining a vector of prediction error, or residuals; $\hat{\varepsilon}^p_{(m)} = Y_i - \hat{Y}_i(0)$.

**End** the loop, collecting $\hat{\boldsymbol{e}}^p = \left\{\hat{\varepsilon}^p_{(1)}, \hat{\varepsilon}^p_{(2)}, ..., \hat{\varepsilon}^p_{(B_1)}\right\}$.

**Step 2.** Apply the GSC method to the original data, obtaining: (1) $\widehat{\text{ATT}}_t$ for all $t > T_0$, (2) estimated coefficients: $\hat{\beta}, \hat{F}, \hat{\Lambda}_{\text{co}}$, and $\hat{\lambda}_{j, j \in \mathcal{T}}$, and (3) the fitted values and residuals of the control units: $\hat{\boldsymbol{Y}}_{\text{co}} = \left\{\hat{Y}_1(0), \hat{Y}_2(0), ..., \hat{Y}_{N_{\text{co}}}(0)\right\}$ and $\hat{\boldsymbol{e}} = \left\{\hat{\varepsilon}_1, \hat{\varepsilon}_2, ..., \hat{\varepsilon}_{N_{\text{co}}}\right\}$.

**Step 3. Start** a bootstrap loop that runs $B_2$ times:

In round $k \in \{1, ..., B_2\}$, construct a bootstrapped sample $S^{(k)}$ by:

$\tilde{Y}_i^{(k)}(0) = \hat{Y}_i(0) + \tilde{\varepsilon}_i, \quad i \in \mathcal{C}$

$\tilde{Y}_j^{(k)}(0) = \hat{Y}_j(0) + \tilde{\varepsilon}^p_j, \quad j \in \mathcal{T}$

in which each vector of $\tilde{\varepsilon}_i$ and $\tilde{\varepsilon}^p_j$ are randomly selected from sets $\hat{\boldsymbol{e}}$ and $\hat{\boldsymbol{e}}^p$, respectively, and $\hat{Y}_i(0) = X_i\hat{\beta} + \hat{F}\hat{\lambda}_i$. Note that the simulated treated counterfactuals do not contain the treatment effect.

Apply the GSC method to $S^{(k)}$ and obtain a new ATT estimate; add $\widehat{\text{ATT}}_{t, t>T_0}$ to it, obtaining the bootstrapped estimate $\widehat{\text{ATT}}^{(k)}_{t, t>T_0}$.

**End** the bootstrap loop.

**Step 4.** Compute the variance of $\widehat{\text{ATT}}_{t, t>T_0}$ using

$$\text{Var}\left(\widehat{\text{ATT}}_t \mid D, X, \Lambda, F\right) = \frac{1}{B} \sum_{k=1}^{B} \left(\widehat{\text{ATT}}^{(k)}_t - \frac{1}{B} \sum_{j=1}^{B} \widehat{\text{ATT}}^{(j)}_t\right)^2$$

and its confidence interval using the conventional percentile method (Efron and Tibshirani 1993).

Algorithm 1: Bootstrap Procedure when `EM=FALSE` (GSC)

The use of out-of-sample predictions is important, as otherwise confidence intervals can be made arbitrarily small by overfitting the factor model. An overfit model will have residuals that are systematically smaller than the true errors. When residuals from this model are permuted to create new bootstrap samples, the resulting bootstrapped estimates will have systematically less variance than the true distribution of estimates implied by the error distribution. In extreme cases where a very large number of latent factors are included in the model, the residuals vanish such that the ATT estimate will have almost zero variance across bootstrap samples, resulting in extreme under-coverage. Out-of-sample errors are not shrunk by overfitting, so the estimated variance will not artificially decrease when prediction errors are used to generate the bootstrap samples, no matter how much the factor model is overfit to the data.

### The Parametric Bootstrap Procedure for IFE-EM

While the parametric bootstrap procedure described in Xu (2017) was designed for the GSC estimator, the `gsynth` package also provided a parametric bootstrap routine for IFE-EM before CRAN version 1.3.1, when this feature was removed without an accompanying explanation in the documentation or release notes we could identify. The most recent version of the package documentation now suggests that the two methods are theoretically incompatible:

> "In early versions, when the estimator is not specified, EM = TRUE activates the Gobillon-Magnac IFE-EM estimator, which has implications for what inferential method is feasible. For example, inference = 'parametric' is not suitable for the IFE-EM and MC estimators given they are super-population based."

We remain uncertain about the conditions under which Xu (2017)'s parametric bootstrap is appropriate for inference in IFE models. We are not aware of theory suggesting that Xu (2017)'s parametric bootstrap would be suitable when researchers use the GSC method to impute counterfactuals, but unsuitable when the IFE-EM method is used for such imputation. For settings with a small number of treated units, the only procedure currently recommended by the `gsynth` package is the original procedure described in Xu (2017).

Setting aside the issue of the parametric bootstrap's suitability for use with the IFE-EM estimator, we investigate the actual software implementation of the parametric bootstrap. Examining the `gsynth` package code prior to version 1.3.1[2], we found that the parametric bootstrap procedure provided for IFE-EM does not use out-of-sample prediction errors when creating the bootstrapped samples. Specifically, step 1 of the previous algorithm is not performed. For complete panels, steps 2-4 are performed drawing all errors from the empirical distribution of the residuals, $\hat{\boldsymbol{e}}$. For unbalanced panels, version 1.2.1 instead generates mean-zero multivariate Normal errors. The GSC implementation uses prediction-error matrices for treated units and residual matrices for controls in this unbalanced-panel case. We summarize the complete-panel IFE-EM procedure below:

BOOTSTRAP PROCEDURE WHEN `EM=TRUE` (IFE-EM)

**Step 1:** Apply the IFE-EM method to the original data, obtaining (1) $\widehat{\text{ATT}}_t$ for all $t > T_0$, (2) estimated coefficients $\hat{\beta}, \hat{F}, \widehat{\lambda_{\text{co}}}$, and $\hat{\lambda}_{j,j\in\mathcal{T}}$, and (3) fitted values and residuals for control units $\hat{\boldsymbol{Y}}_{\textbf{co}} = \left\{\hat{Y}_1(0), \hat{Y}_2(0), ..., \hat{Y}_{N_{\text{co}}}(0)\right\}$ and $\hat{\boldsymbol{e}} = \left\{\hat{\varepsilon}_1, \hat{\varepsilon}_2, ..., \hat{\varepsilon}_{N_{\text{co}}}\right\}$

**Step 2:** Start a bootstrap loop that runs $B_2$ times:

  In round $k \in \{1, ..., B_2\}$, construct a bootstrapped sample $S^k$ by:

    $\tilde{Y}_i^{(k)} = \hat{Y}_i(0) + \tilde{\varepsilon}_i$

  In which each vector of $\hat{\varepsilon}_i$ is sampled from the set of in-sample residuals $\hat{\boldsymbol{e}}$, and $\hat{Y}_i(0) = X_i\hat{\beta} + \hat{F}\hat{\lambda}_i$

  Apply the IFE-EM method to $S^k$ and obtain a new ATT estimate; add $\widehat{\text{ATT}}_{t,t>T_0}$ to it, obtaining the bootstrapped estimate $\widehat{\text{ATT}}^k_{t,t>T_0}$

  **End** the bootstrap loop.

**Step 3:** Compute a confidence interval for $\widehat{\text{ATT}}_{t,t>T_0}$ using the percentile method[3]

Algorithm 2: Bootstrap Procedure when `EM=TRUE` (IFE-EM)

This differs from the bootstrap procedure described in Algorithm 1, as it does not use out-of-sample prediction errors in creating new bootstrapped samples. Accordingly, this procedure can generate undersized confidence intervals when the factor model is overfit.

In `gsynth` 1.2.1, two further features of the GSC bootstrap depart from the published pseudocode: the final bootstrap resamples control identities and their fitted outcomes and design variables; the prediction-error stage repeats a selected control $N_{\text{tr}}$ times as pseudo-treated units, where $N_{\text{tr}}$ is the number of treated units. We found no prose explanation of either feature in the contemporary documentation and public materials we searched. Our corrected IFE-EM bootstrap inherits both features from that GSC code when restoring prediction-error estimation. We do not investigate them further.

[2]The `gsynth` code that performs the parametric bootstrap when IFE-EM is used can be found from lines 1385-1420 of `R/gsynth.R` in the original version of `gsynth` uploaded to GitHub (commit `8d2997d`). This code was eventually moved to lines 2687-2760 of `R/core.R` in later versions of the package (commit `e4525ad`), and was removed prior to `v1.3.1` of the package, which involved substantial refactoring of the internal architecture of the package.

[3]In version 1.1.7, this was switched to a Normal approximation to guarantee symmetry of CIs about the point estimate.

For unbalanced panels, our corrected bootstrap draws residuals from the same multivariate normal distribution used by the GSC bootstrap.

# A Toy Example

We construct a simulated dataset to show how this overfitting issue can lead to underestimation of standard errors. To demonstrate this phenomenon, we construct a dataset consisting of 40 treated units and 200 control units. Units are observed over 80 time periods, with treatment beginning at $t = 41$. In these simulations, outcomes are purely noise such that $Y_{i,t} = \varepsilon_{i,t}$. Errors are independent between units, but correlated within units following a multivariate normal distribution:

$$\varepsilon_i \sim N(0, K)$$
$$K_{t,t'} = \begin{cases} 4.01 & \text{if } t = t' \\ 4e^{-\frac{(t-t')^2}{300}} & \text{if } t \neq t' \end{cases}$$

We illustrate with a single draw from this data generating process. Panel (a) illustrates the data, demonstrating the long-term correlation of the errors in this setting. Never-treated units are shaded in light grey, while treated units are orange in the pre-treatment period and highlighted in red after treatment turns on.

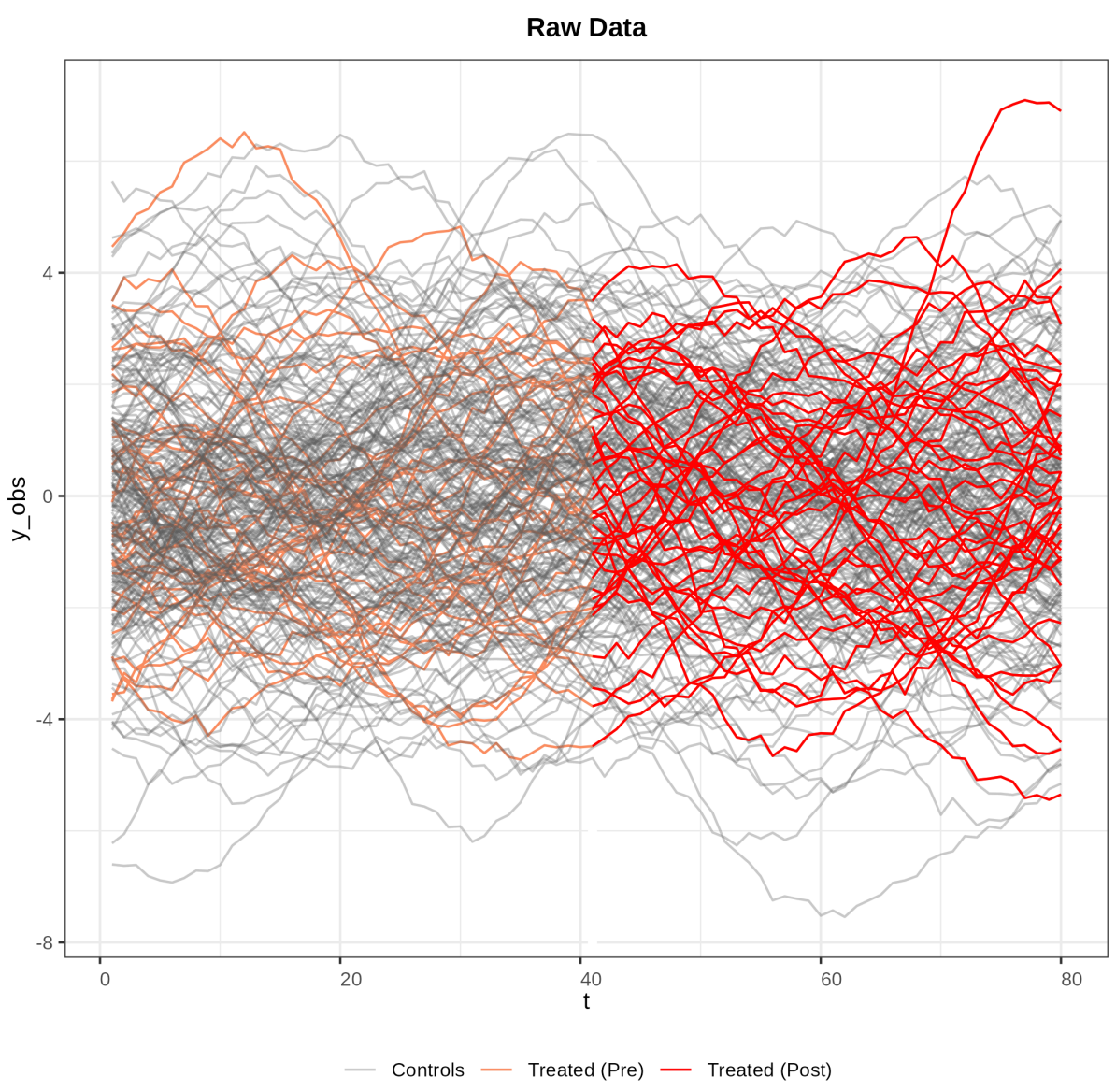

a) Synthetic Data (No Treatment Effect)

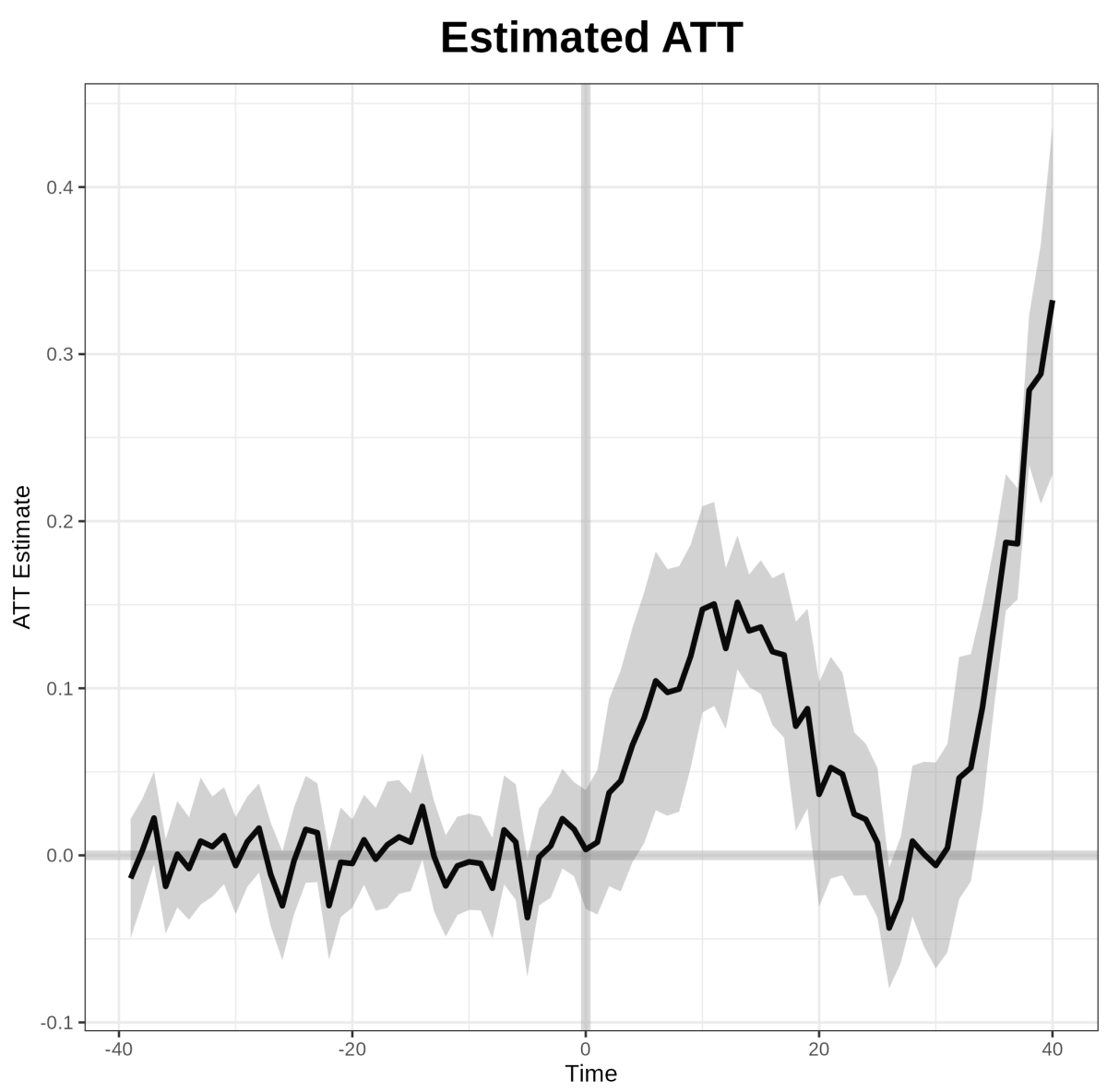

b) ATT Estimate Produced using IFE-EM + Parametric Bootstrap (gsynth 1.2.1)

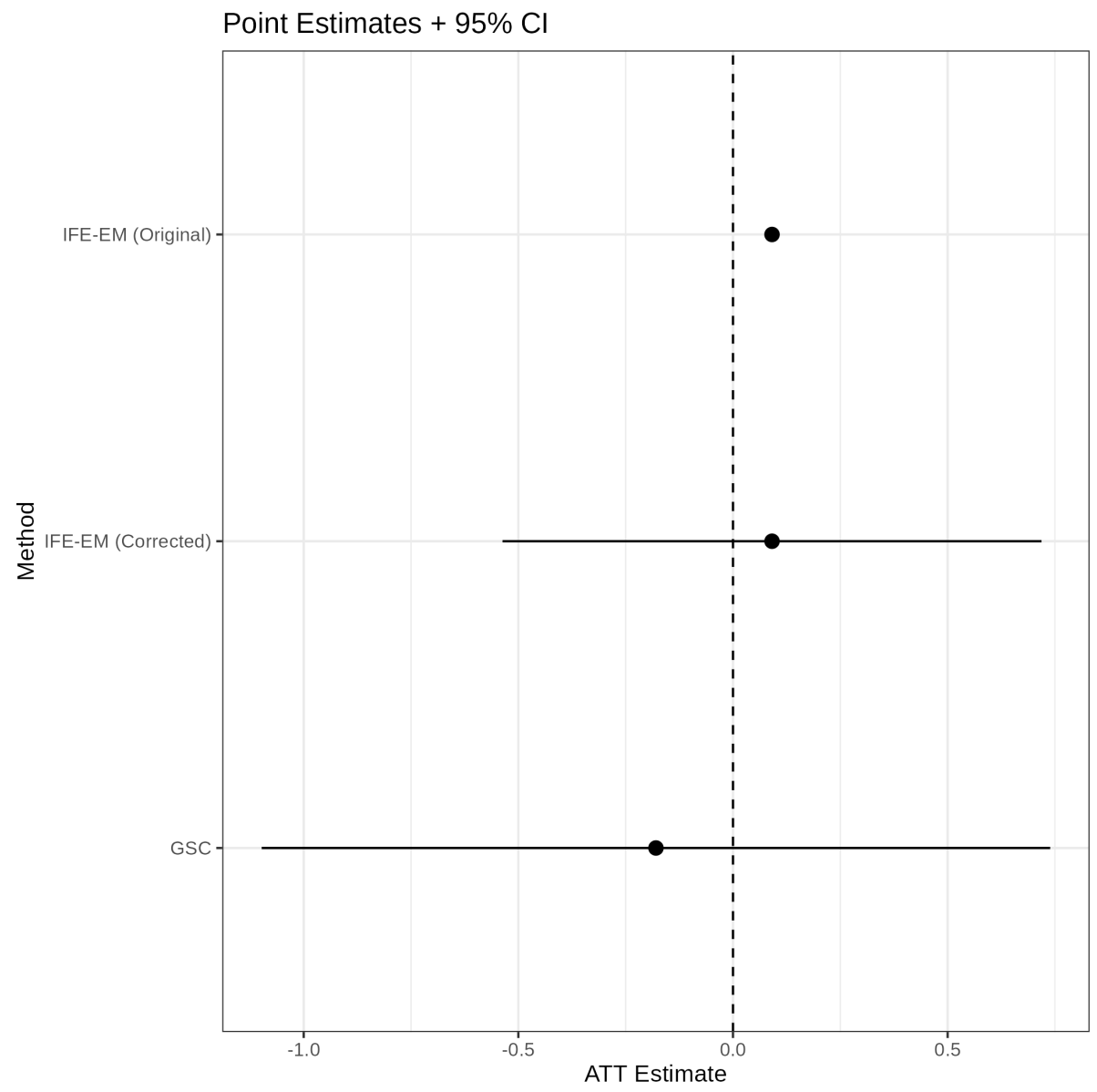

c) Estimates with Confidence Intervals

Panel (b) of the figure shows a treatment effect plot produced by `gsynth` using IFE-EM. After treatment turns on, the method identifies a strong and highly significant effect which fluctuates wildly across post-treatment time periods. The average treatment effect is estimated as $\widehat{\text{ATT}} = 0.09$ with a standard error of 0.006 implying almost complete certainty in the estimate.

The serial correlation in the error terms encourages the model to overfit; examining the `gsynth` model output, we see that the cross-validation procedure has selected 7 factors despite the fact that the true DGP includes zero latent factors. Overfitting occurs because the cross-validation procedure selects the number of factors that minimizes prediction error on a left-out time period but under serial correlation, errors in the left-out period are not independent of the adjacent periods. This means that the observed error in one time period can successfully be used to predict the left-out observation, rewarding overfitting.

The 7 factors fit by the model obtain an R-squared value of .997 within the control group, meaning the parameters of the model explain 99.7% of the variation in the outcomes, with the residuals accounting for the remaining 0.3% of the variation. The bootstrapped estimates that come from resampling the remaining 0.3% of the outcome are essentially identical between samples, resulting in minuscule confidence intervals.

We note that these minuscule confidence intervals only happen if the model is fit using the parametric bootstrap procedure that does not use out-of-sample prediction errors. Panel (c) of the figure shows the relationship between the estimates produced by the package with and without expectation-maximization, as well as with a version of expectation maximization that uses a corrected version of the parametric bootstrap which restores the prediction-error step described in Algorithm 1. The confidence intervals from the GSC and IFE-EM + corrected bootstrap variants cover the true value of zero, and are of somewhat similar width, while the confidence interval from IFE-EM as implemented in the `gsynth` package is too small to be visible.

## Simulation Study: Correlates of State Policy Project

We have documented the potential for overfitting in a stylized setting, but is this phenomenon a concern when analyzing real datasets? In this section, we perform a simulation study using state-level economic data from the Correlates of State Policy Project (Grossmann, Jordan, and McCrain 2021) in which we randomly assign cohorts of states to be 'treated' and calculate the rate at which the `gsynth` package's implementation of IFE-EM and parametric bootstrap identifies significant effects. We show that the false positive rate can reach nearly 100% for some outcomes. Averaging over all simulations, coverage is 49.1%.

We choose a state-level panel simulation because the `gsynth` package is most commonly applied to estimate the effect of an exogenous policy shock on the economic or political trajectories of a set of treated states (Gilens, Patterson, and Haines 2021; Grumbach 2022; Xu 2017).

To evaluate the coverage of IFE-EM with the parametric bootstrap, we perform the following steps: First, we download the Core Academic Release of the Correlates of State Policy Project dataset (Grossmann, Jordan, and McCrain 2021), which contains year-by-state measures of 66 public opinion, policy, and economic outcomes. We retain outcomes that are observed over more than 12 years, feature a missing rate less than 2%, and have more than 300 distinct values of the outcome.

These filtering steps leave us with a set of 16 outcomes, listed in the Appendix. These outcomes include measures of public opinion towards spending on social issues (the environment, and defense spending), as well as economic and demographic outcomes such as the Gini coefficient in each state, and the state's union density. The original code retains years strictly between the last observed year minus 20 and the last observed year, leaving $T = 19$ for every panel. The population, income, GDP per capita, Gini coefficient, violent crime, and policy mood panels include the District of Columbia in addition to the 50 states; the other ten panels contain 50 states.

For each outcome, we assess the coverage of the `gsynth` 1.2.1 implementation of IFE-EM with the parametric bootstrap by randomly designating 5 units as treated at time $t = 12$, leaving 11 pre-treatment observations and 8 treatment observations for each treated unit. We record the point estimate and the associated confidence interval. We repeat the treatment assignment step 200 times such that we can estimate the coverage of the method for each outcome. As a point of comparison, we also perform the same exercise using TWFE with standard errors clustered at the state level, and the GSC with the parametric bootstrap also implemented in version 1.2.1 of `gsynth`. (We omit a simulation with our corrected parametric bootstrap procedure due to a long expected runtime). The results of this simulation study are shown below:

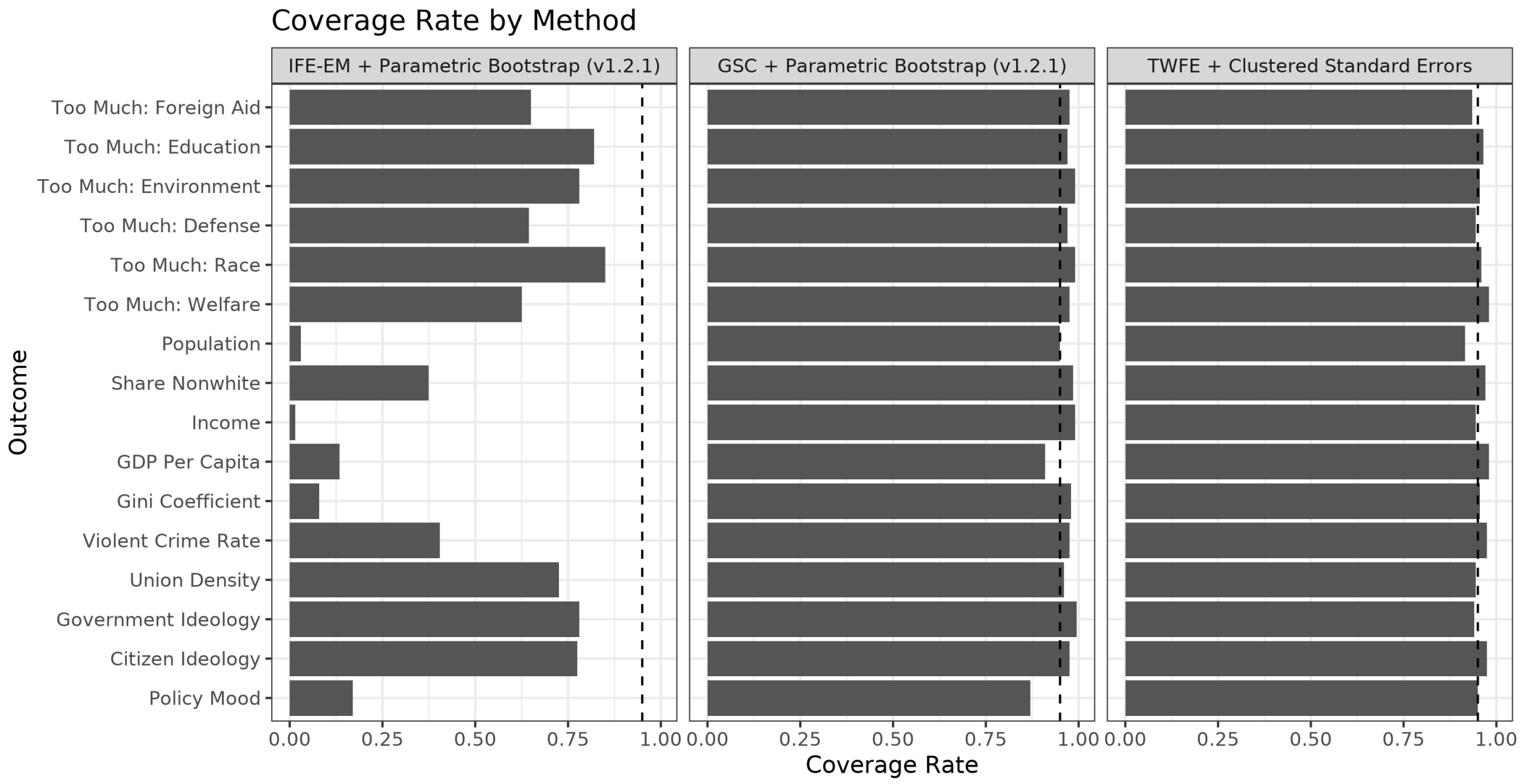

We find that for each of the 16 outcomes under study, the package yields a coverage rate that is significantly lower than 95%. For 7 outcomes, the coverage rate is significantly lower than 50%, suggesting that the method is more likely than not to identify the effect of a placebo treatment as significant in these cases. The two alternative methods appear to perform roughly as expected in this setting: the 95% confidence intervals produced by TWFE obtain above 90% coverage across all outcomes, while intervals produced by the GSC + Parametric bootstrap procedure obtain coverage above 90% across 15 of the 16 outcomes.

# Reanalysis of Affected APSR Papers

We have demonstrated in simulation studies that using IFE-EM and the parametric bootstrap procedure in `gsynth` prior to version 1.3.1 would severely underestimate standard errors. How should these results shape our understanding of papers that have used `gsynth`? We identify and reanalyze three papers published in the APSR that use this combination of parameters.

To identify papers that use `gsynth` with this combination of parameters, we scrape the APSR dataverse to obtain a list of every paper that uses the `gsynth` library as part of its replication archive. We then subset these papers to consider only analyses that use both IFE-EM and the parametric bootstrap. We cross-check using a list of all papers that cite the original GSC article (Xu 2017) via CrossRef.

We find four papers that use this combination of parameters (Alsaadi 2025; Blair, Grossman, and Weinstein 2021; Eibl and Hertog 2023; Gilens, Patterson, and Haines 2021). Of these four papers, we find one paper (Blair, Grossman, and Weinstein 2021) that fits the estimator using zero latent factors, in which case the estimator is logically equivalent to using a TWFE model to impute counterfactual observations, and is not plausibly affected by overfitting. We omit this paper from our analysis, leaving three papers (Alsaadi 2025; Eibl and Hertog 2023; Gilens, Patterson, and Haines 2021), which we reanalyze below.

We will report three estimates and confidence intervals for each finding:

- We produce a pure replication of the estimate and 95% CI in the paper using IFE-EM and the parametric bootstrap as implemented in `gsynth` 1.2.1.
- We use the same IFE-EM estimate as above, but use our corrected implementation of Xu (2017)'s parametric bootstrap to construct CIs.

- We report estimates using the GSC method and parametric bootstrap, as proposed in Xu (2017).

Our use of the GSC method and parametric bootstrap should not be taken as an endorsement of the approach. It is important to report regardless. This approach corresponds to the current recommendations of the `gsynth` package, and thus serves as a benchmark representing current practice. And all three papers under reanalysis describe their inferential strategy as Xu (2017)'s GSC method, and thus this strategy matches the textual description of the methodology.

## Reanalysis of Gilens, Patterson, and Haines (2021)

Gilens, Patterson, and Haines (2021) employ IFE-EM to estimate the effects of the *Citizens United* decision on the passage of corporate-friendly legislation. The underlying logic is that states that did not have an existing ban on corporate campaign spending were 'untreated' by the decision, and can serve as control units for states which had bans invalidated by the decision. Gilens, Patterson, and Haines (2021) find that the *Citizens United* Supreme Court decision led to lower corporate tax rates and tort-law policies that were more beneficial to corporations. Their main corporate-tax analysis uses a state-level panel of 42 states over 17 time periods (10 pre-treatment and 7 post-treatment). The authors also analyze the effects of the decision on three policies that less directly affect corporations - abortion laws, eminent domain laws, and gun control, and find no evidence for effects of the decision on these outcomes. Some of the largest effects were observed for corporate tax rates, where the authors find that *Citizens United* lowered corporate tax rates by approximately 4 percent across affected states (Gilens, Patterson, and Haines 2021, p 1078). Among states that had bans on corporate but not union campaign spending, the effects are larger; these states saw an average decrease in the corporate tax rate of 8 percent (Gilens, Patterson, and Haines 2021, p 1078).

We reproduce the authors' analyses in the figure below, which displays the estimated effect of the *Citizens United* decision on the five main outcomes analyzed in the paper. We reanalyze the panels using the GSC method and IFE-EM with the corrected parametric bootstrap procedure. We follow the authors' choice of using a scaled measure of corporate tax rates, such that a 4 percent reduction in corporate tax rates corresponds to the 2.8-point unit decrease reported in the figure below.

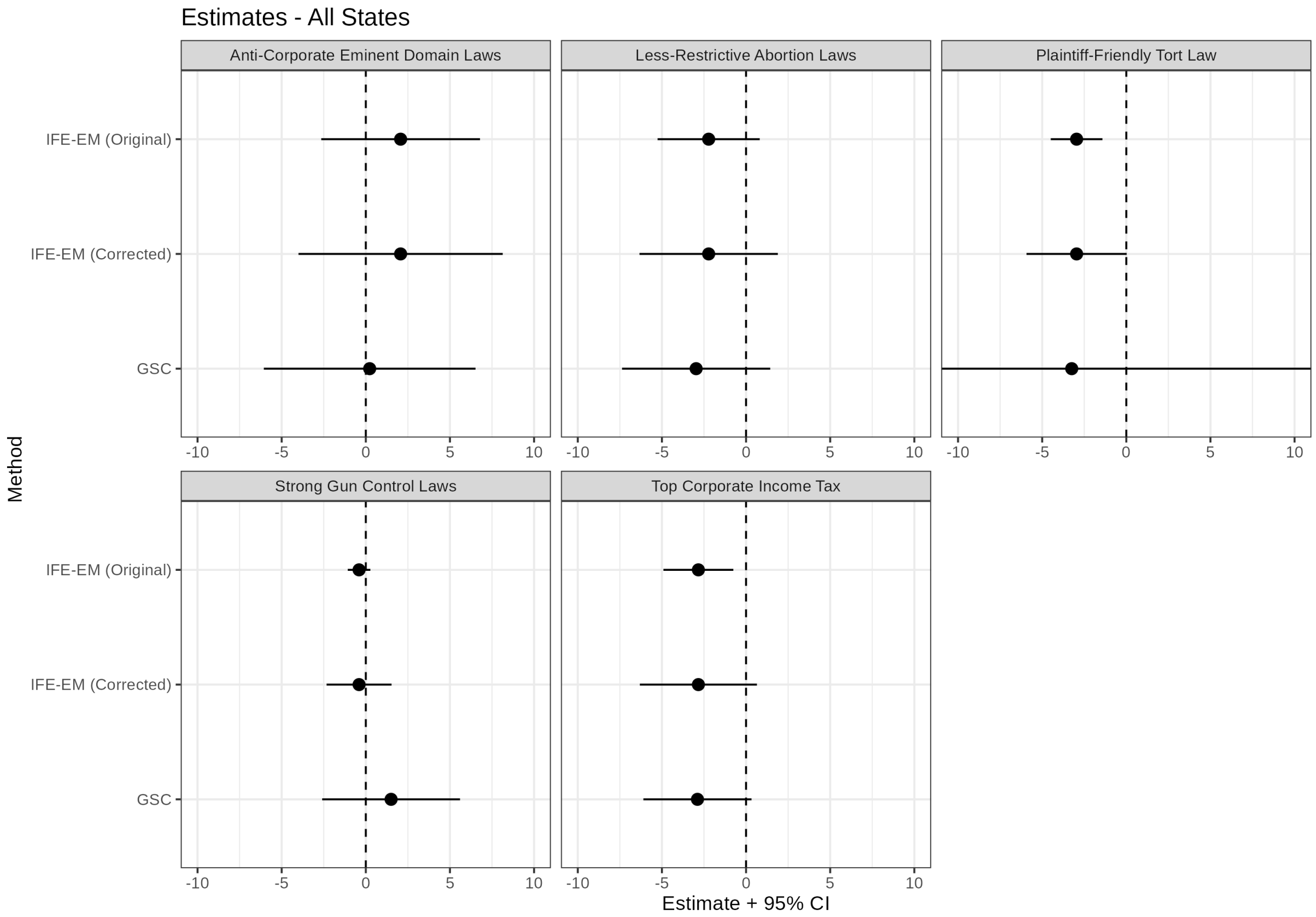

Figure 3: Estimates of Effect of *Citizens United* on Policy Outcomes. Replication of Table 2 from Gilens, Patterson, and Haines (2021)

We show that the standard errors were greatly underestimated. For the primary outcome reported, Top Corporate Income Tax Rate, correcting the parametric bootstrap increases the SE by 67%. Switching to GSC has a similar consequence, though other outcomes can be more dramatically affected by the change.[4] Across all outcomes, we find that the estimated effects of the *Citizens United* Decision are rendered statistically insignificant at the 95% level when reanalyzed using either the corrected parametric bootstrap or with the GSC method.

The authors also perform the same analysis on a subset of states that had a ban on corporate campaign but not union campaign spending, finding larger effects. We replicate these analyses in the Appendix. We find that correcting the parametric bootstrap increases the estimated SE for the estimated effect on Top Corporate Income Tax Rate by 794% in this subgroup. All findings lose statistical significance when reanalyzed with either the corrected parametric bootstrap or with the GSC method.

## Reanalysis of Alsaadi (2025)

Alsaadi (2025) uses IFE-EM as part of 5 quantitative analyses and a qualitative case study to investigate whether minority autocracies that exclude a single majority ethnic group are able to instigate activism to sustain their regime more effectively than comparable autocracies that exclude minority ethnic groups. The underlying causal theory is that members of the ethnic minority ruling group are forced to almost unconditionally support the autocratic regime, as if the regime falls, they will likely be subject to rule by the majority ethnic group. This theory predicts that the onset of a minority autocracy that excludes a single majority ethnic group will increase the intensity of mobilization in support of the autocracy, while no such change will be observable when a regime that excludes multiple ethnic groups comes into power.

[4]We note that the range of these plots is trimmed to $[-10,10]$: The confidence interval produced by the GSC method for the tort law outcome is infinitely wide because the outcome has essentially zero variation in post-treatment periods among never-treated units.

Consistent with this theory, Alsaadi (2025) finds that a transition to a minority autocracy that excludes a single majority ethnic group is associated with a statistically significant increase in mobilizations in support of autocracy. In contrast, a transition to a minority autocratic regime that excludes other minorities is not associated with any detectable change in the amount of mobilization for autocracy. Both analyses are conducted on panels of countries featuring approximately 120 countries with 9 countries in the 'minority excluding single majority' treatment arm, and 17 countries in the 'minority excluding minorities' group. The data are yearly observations from 1900-2015 for a total of 116 time periods. We reproduce the results of this analysis below, also reporting the results with the corrected parametric bootstrap and using the GSC method:

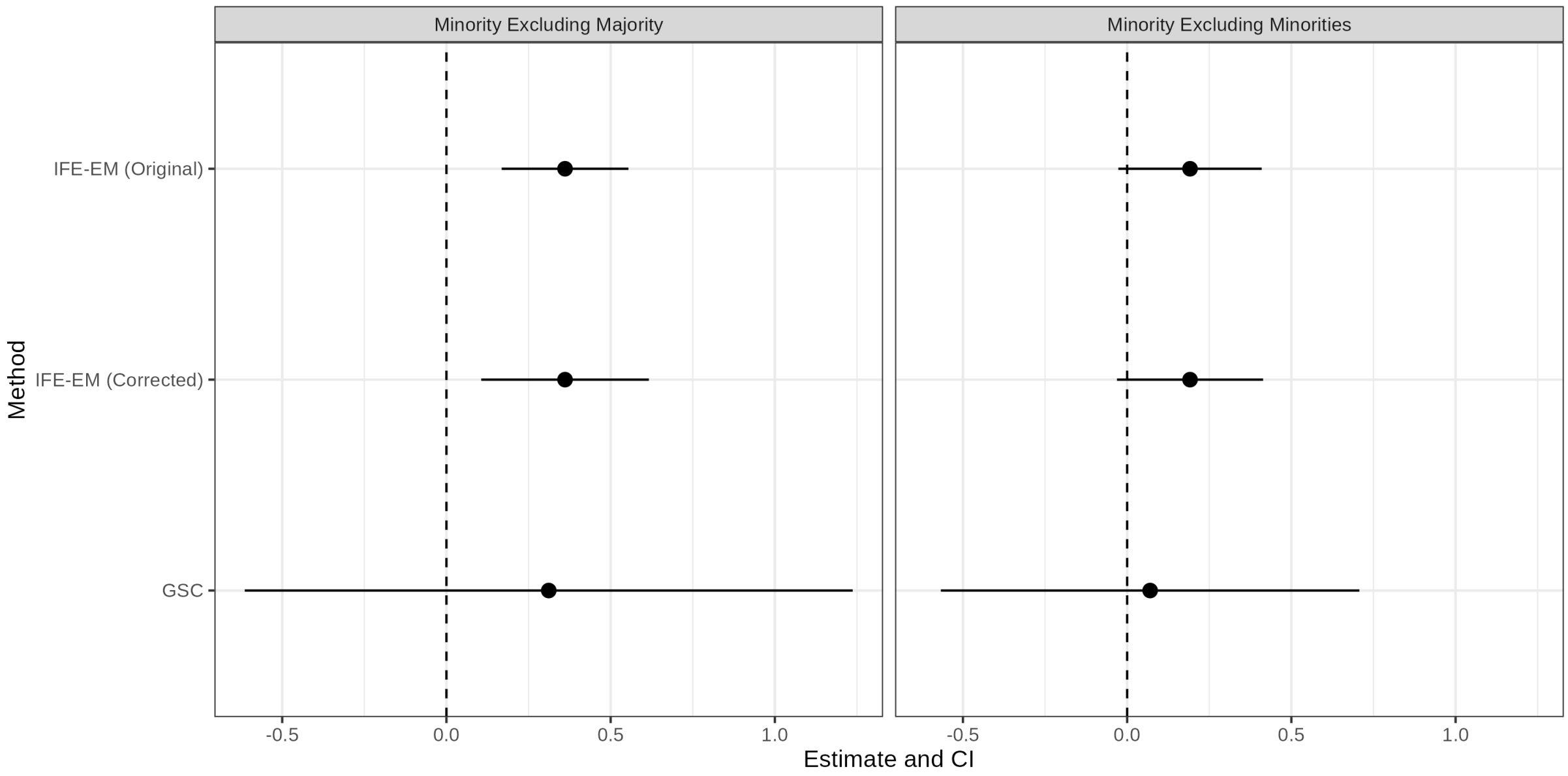

Figure 4: Reanalysis of Alsaadi (2025) Table 4 Estimates

The first finding, that the onset of a minority regime that excludes a single majority is associated with an increase in mobilization, remains statistically significant using the corrected parametric bootstrap procedure, but the estimated SE is 28% larger. Use of the GSC procedure has more dramatic consequences, losing significance, and seeing the SE become 392% larger. The second finding, the (null) effect of transitioning to a minority regime that excludes multiple ethnic groups, remains insignificant across the two alternative specifications. The SE on this estimate is largely unaffected by correcting the parametric bootstrap (a 2% increase), though use of the GSC procedure increases the standard errors by 201%.

## Reanalysis of Eibl and Hertog (2023)

Eibl and Hertog (2023) estimate the effects of anti-regime popular movements with broad public support (subversions) within oil-rich countries on the provision of public goods, specifically educational equality, health equality and primary and secondary school enrollment. The authors' theory predicts that *center-seeking subversions* or mobilizations that aim to change the ruling regime will lead to an increase in the provision of public services in oil-rich countries, as rulers share oil rents in an effort to quell protests.

Consistent with this theory, the authors find that the onset of subversion is associated with an increase in state-provided public goods in oil-rich, but not oil-poor states. The outcome data span 1900-2018 (119 years), while the fitted panels contain 61–65 time periods and 73–152 countries, depending on the outcome and subversion group.

We replicate the point estimates from Table 2 of the paper. The original paper reported the point estimates and CIs associated with the IFE-EM estimator and parametric bootstrap as implemented in `gsynth` 1.2.1, for the ATT in the 15th year following subversions. In the figure below, we show the original estimates, as well as the estimates under reanalysis with the corrected parametric bootstrap procedure, and the GSC method.

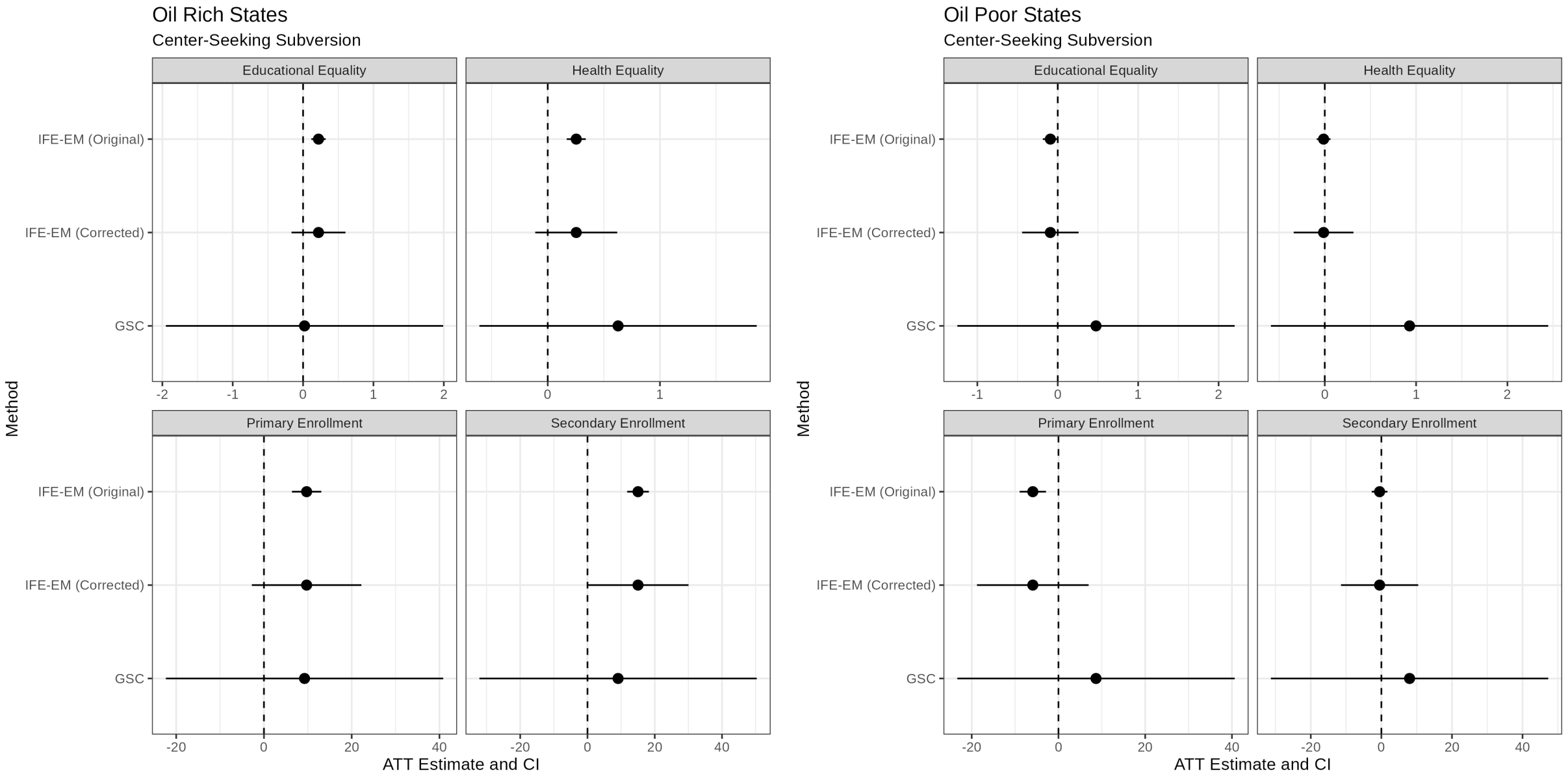

Figure 5: Reanalysis of effects of center-seeking subversions analysis from Eibl and Hertog (2023)

In general, we see a marked increase in uncertainty after correction. Depending on the outcome, correcting the parametric bootstrap causes the SEs to increase by anywhere from 252% to 402%. Using GSC causes SEs to increase in size by 790% to 2498%.

The main finding from the original paper, that a center-seeking subversion leads to increased provision of public goods in oil-rich countries, largely does not survive reanalysis. While the original analysis identifies a significant effect of a center-seeking subversion across four outcomes; only one outcome (Secondary Enrollment) remains significant ($p = 0.05$) after correcting the parametric bootstrap. None remain significant using GSC.

Turning to oil-poor states, the IFE-EM method identifies significant effects across two outcomes (Education Equality and Primary Enrollment), but neither of these effects is significant when the GSC algorithm and associated parametric bootstrap procedure are used. These results are consistent with the authors' theory which does not imply any predictable relationship between subversions and public good provision in oil-poor states, although we note that the wide confidence intervals mean that even substantively large effects cannot be ruled out.

The authors also perform the same analysis on a panel of states that underwent separatist subversions, finding that while separatist subversions appear to have an effect on some outcomes, there are no predictable relationships between separatist subversions and public goods provisions. We replicate this analysis in the Appendix using the corrected parametric bootstrap and the GSC method and find that all effects are no longer significant using either the GSC or corrected IFE-EM procedure. This suggests that any associations are compatible with chance fluctuations in the outcomes, but we note that the confidence intervals are so wide as to be compatible with even large effects.

## Conclusion

In this note, we have demonstrated that when two estimation options provided by the `gsynth` package (parametric inference and the `IFE-EM` estimator) were used in tandem, the resulting confidence intervals could be substantially undersized, in some cases by orders of magnitude. While this issue does not affect analyses performed using the most recent versions of the package, we believe knowledge of this issue should spur a re-evaluation of papers that have used the method. Examining only papers published in the *American Political Science Review*, we find three papers that require correction as a result. In the appendix we provide a guide to assessing whether or not the implementation error in `gsynth` has affected their analysis.

## Declaration of generative AI and AI-assisted technologies

During preparation of this manuscript, the authors used ChatGPT (GPT 5.5 and 6) to assist with drafting, editing and checking arguments. The authors reviewed and edited the content as necessary and take full responsibility for the content of the manuscript.

## Timeline of Key Events

- **July 2016** - Gobillon and Magnac (2016) propose an interactive fixed-effects estimator fit with Expectation Maximization.
- **Feb 2017** - Xu (2017) proposes the Generalized Synthetic Control Method and associated parametric bootstrap procedure.
- **March 2017** `gsynth` version 1.0.2 is released to CRAN, supports both the GSC estimator and the IFE-EM estimator. Both methods are colloquially called "GSC". The parametric bootstrap is the default inference procedure. Its implementation, when the IFE-EM estimator is used, does not match the algorithm in Xu (2017).
- **Feb 2022** Vignette is published to the `gsynth` package website that demonstrates the use of IFE-EM with the parametric bootstrap.
- **Dec 2025** The option to use the parametric bootstrap and IFE-EM together is removed following a major refactor to the `gsynth` codebase (release 1.3.1) such that it is now largely a wrapper of the `fect` package.
- **March 2026** We discover the implementation error. We correspond with Xu regarding `gsynth`'s behavior in versions prior to `1.3.1` when the parametric bootstrap and `IFE-EM` are used.
- **March 2026** The `gsynth` package documentation is updated to indicate that the parametric bootstrap is "not suitable" for IFE-EM estimator due to a theoretical incompatibility, and acknowledges an "erroneous implementation" of the algorithm used prior to 1.3.1 when `EM=T` and `inference=parametric`.

Listing 1: Timeline of Key Events

# Guide to Assessing Affected Papers

We recommend using the following procedure to assess whether the results in a published paper are affected.

## Step 1:

Was the version of `gsynth` greater than or equal to 1.3.1[5]? If so, this error cannot occur. No reanalysis is necessary.

## Step 2:

If the version of `gsynth` used was less than 1.3.1, check all uses of the `gsynth` command for the parametric bootstrap with the EM algorithm. The diagram below gives examples of four code snippets using the `gsynth` command and classifies them by vulnerability to the issue highlighted in this note.

```r
gsynth_out <- gsynth(
  outcome ~ treatment,
  index = c("unit", "time"),
  data = data,
  se = TRUE,
  inference = "parametric",
  EM = TRUE
)
```

a) Affected by issue (IFE-EM + parametric bootstrap)

```r
gsynth_out <- gsynth(
  outcome ~ treatment,
  index = c("unit", "time"),
  data = data,
  se = TRUE,
  inference = "parametric",
  EM = FALSE
)
```

b) Not affected by issue (GSC + parametric bootstrap)

```r
gsynth_out <- gsynth(
  outcome ~ treatment,
  index = c("unit", "time"),
  data = data,
  se = TRUE ,
  inference = "nonparametric",
  EM = TRUE
)
```

c) Not affected by issue (IFE-EM and nonparametric bootstrap)

```r
gsynth_out <- gsynth(
  outcome ~ treatment,
  index = c("unit", "time"),
  data = data,
  se = TRUE ,
  estimator = "mc"
)
```

d) Not affected by issue (Matrix Completion estimator)

[5] Version 1.3.1 was released to CRAN in December 2025.

## Outcomes in Empirical Monte Carlo Exercise

The table below describes each of the 16 outcomes used in our empirical Monte Carlo Exercise. Descriptions are reproduced from the Correlates of State Policy Project codebook (Grossmann, Jordan, and McCrain 2021).

| Outcome Number | Name | Human-Readable Name | Description |
|---|---|---|---|
| 1 | `anti_aid` | Too Much: Foreign Aid | Estimated proportion of citizenry who believe we are spending 'too much' on foreign aid (note: proportion for state-year is estimated by national sample from the General Social Survey and multilevel regression and post-stratification) |
| 2 | `anti_education` | Too Much: Education | Estimated proportion of citizenry who believe we are spending 'too much' on education (note: proportion for state-year is estimated by national sample from the General Social Survey and multilevel regression and post-stratification) |
| 3 | `anti_environment` | Too Much: Environment | Estimated proportion of citizenry who believe we are spending 'too much' on the environment (note: proportion for state-year is estimated by national sample from the General Social Survey and multilevel regression and post-stratification) |
| 4 | `anti_defense` | Too Much: Defense | Estimated proportion of citizenry who believe we are spending 'too much' on defense (note: proportion for state-year is estimated by national sample from the General Social Survey and multilevel regression and post-stratification) |
| 5 | `anti_race` | Too Much: Race | Estimated proportion of citizenry who believe we are spending 'too much' on assistance to blacks (note: proportion for state-year is |
| | | | estimated by national sample from the General Social Survey and multilevel regression and post-stratification) |
| 6 | `anti_welfare` | Too Much: Welfare | Estimated proportion of citizenry who believe we are spending 'too much' on welfare for the poor (note: proportion for state-year is estimated by national sample from the General Social Survey and multilevel regression and post-stratification) |
| 7 | `poptotal` | Population | Total population per state |
| 8 | `nonwhite` | Share Nonwhite | Proportion of the state population that is nonwhite |
| 9 | `incomepcap` | Income | Per capita personal income is the income that is received by persons from all sources. Total personal income divided by total midyear population. |
| 10 | `gsppcap` | GDP Per Capita | Per capita GDP by state is the state counterpart of the nation's gross domestic product (GDP) divided by the residents of the state. Current dollars per state resident. Per capita GDP by state is derived as the sum of the GDP originating in all the industries in a state divided by the population of the state. |
| 11 | `gini_coef` | Gini Coefficient | State income inequality measured by Gini coefficient |
| 12 | `vcrimerate` | Violent Crime Rate | Estimated violent crime rate by state. Violent crimes include murder and nonnegligent manslaughter, forcible rape, robbery, and aggravated assault. |
| 13 | `union_density` | Union Density | Proportion of (non-agricultural) workforce represented by a union |
| 14 | `inst6014_nom` | Government Ideology | This was the authors' second measure of state government ideology. Instead of relying on ADA and COPE scores to construct a measure, the authors rely on 'Common-Space' congressional ideology scores to construct their measure of state party ideology (available at: http://voteview.com/basic.htm) |
| 15 | `citi6013` | Citizen Ideology | Measure of citizen ideology, from liberal to conservative. Higher values indicate more liberal. |
| 16 | `mood` | Policy Mood | An over-time, state-level measure of Stimson's (1999) policy mood |

# Reanalysis of Gilens, Patterson, and Haines (2021), States with Ban on Corporate Spending Only

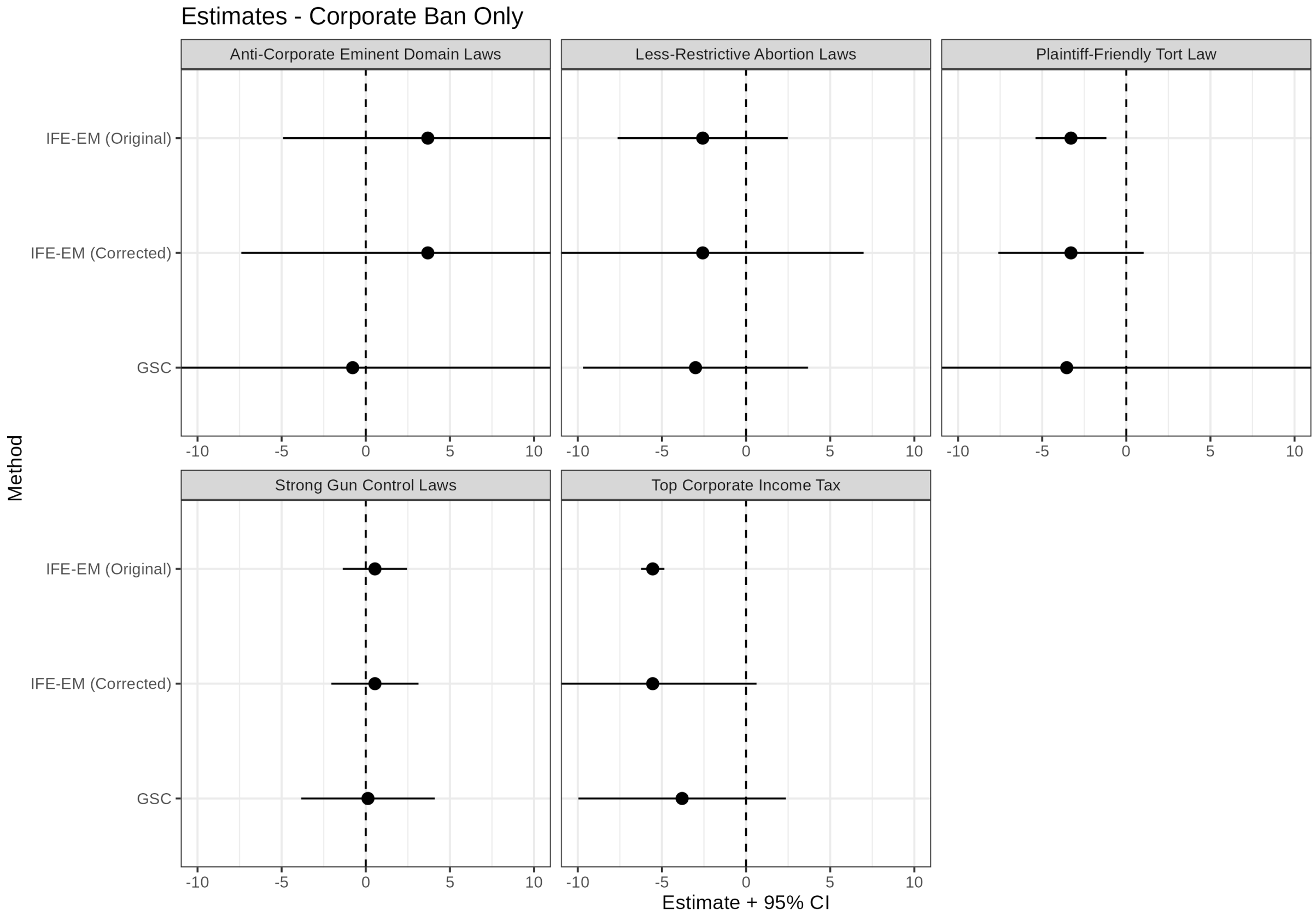

Figure 6: Estimates of Effect of *Citizens United* on Treated States With Preexisting Bans on Corporate but not Union Spending. Replication of Table 2 from Gilens, Patterson, and Haines (2021)

# Reanalysis of Eibl and Hertog (2023), Separatist-Seeking Subversions

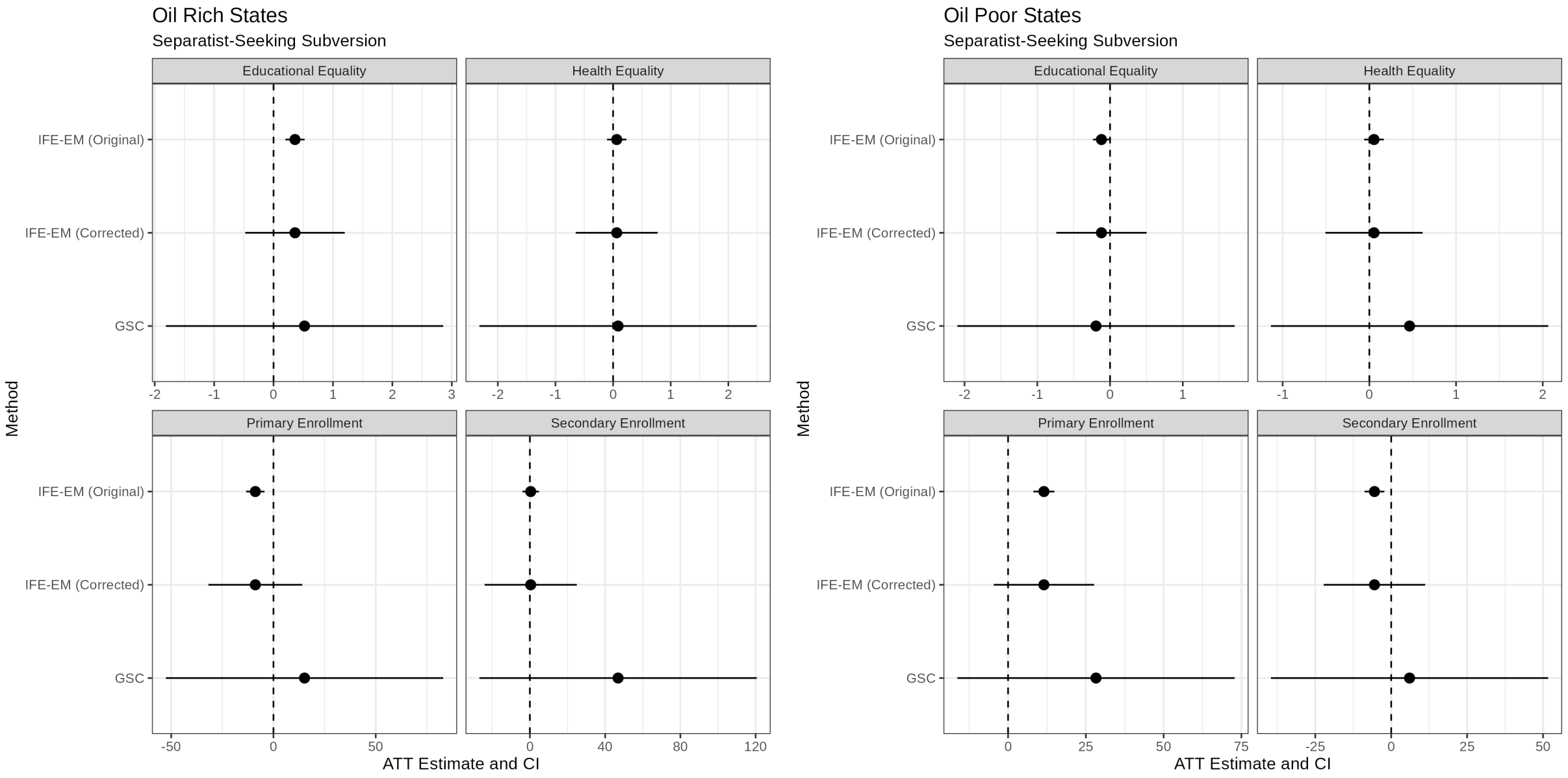

Figure 7: Reanalysis of Effects of Separatist-Seeking Subversions Analysis from Eibl and Hertog (2023)